\documentclass[aps,pre,amsmath,amssymb]{revtex4-2}

\usepackage{polski}
\usepackage[english]{babel}
\usepackage{bm}
\usepackage{graphicx}
\usepackage{amsmath}
\usepackage{mathrsfs}

\begin{document}

\newcommand{\uu}[1]{\underline{#1}}
\newcommand{\pp}[1]{\phantom{#1}}
\newcommand{\be}{\begin{eqnarray}}
\newcommand{\ee}{\end{eqnarray}}
\newcommand{\ve}{\varepsilon}
\newcommand{\vp}{\varphi}
\newcommand{\vs}{\varsigma}
\newcommand{\Tr}{{\,\rm Tr\,}}
\newcommand{\Trr}{{\,\rm Tr}}
\newcommand{\pol}{\frac{1}{2}}
\newcommand{\sgn}{{\rm sgn}}
\newcommand{\Mo}{\mho}
\newcommand{\Om}{\Omega}

\title{Acceleration  deforms exponential decays into generalized Zipf-Mandelbrot laws}
\author{Marek Czachor}
\affiliation{
Instytut Fizyki i Informatyki Stosowanej,
Politechnika Gdańska, 80-233 Gdańsk, Poland
}

\begin{abstract}
An exponentially decaying system looks as if its decay was a generalized power or double-exponential law, provided one takes into account the relativistic time dilation in a detector, the delay of the emitted signal, and the accelerations of both the source and the detector. The same mathematical formula can be found in generalizations of the Zipf-Mandelbrot law in quantitative linguistics and in the dynamics of ligand binding in heme proteins. The effect is purely kinematic and is not related to the various dynamic phenomena that can accompany accelerated motion of sources or detectors. The procedure used can also be seen as a form of clock synchronization near an event horizon. \end{abstract}

\maketitle

\section{Watching a decay}

\begin{figure}
\includegraphics[width=4.5 cm]{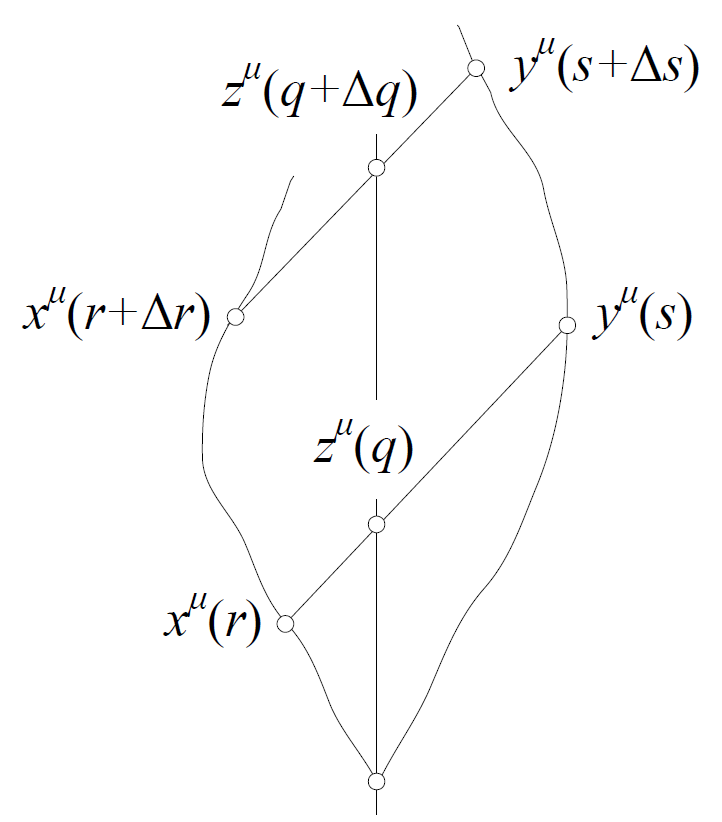}
\includegraphics[width=5.5 cm]{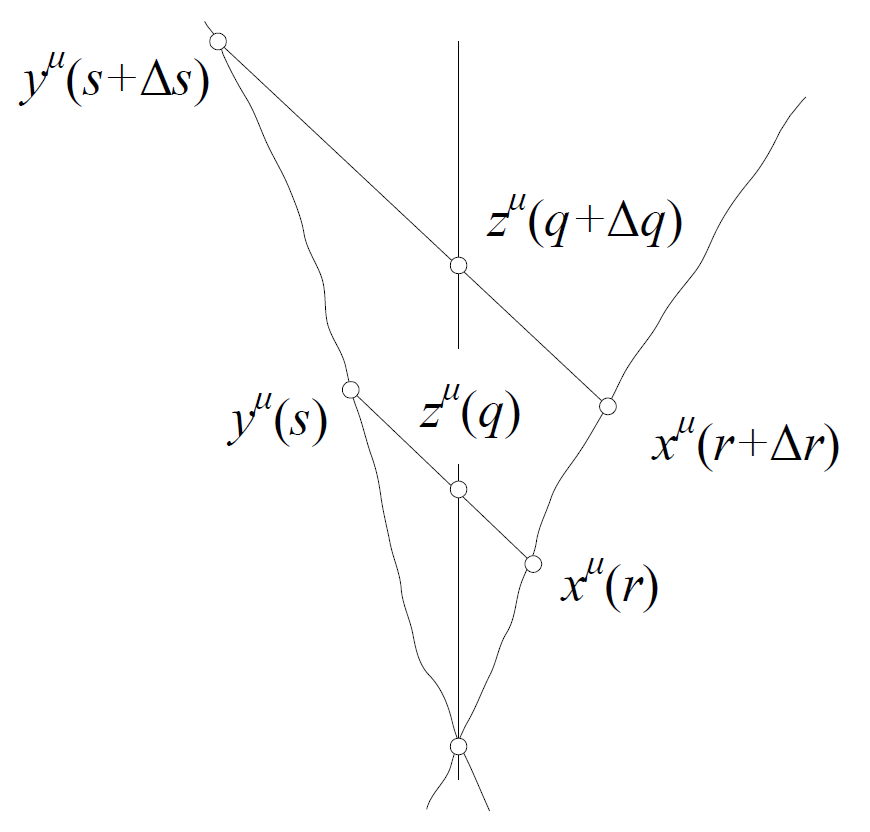}
\caption{
The geometry of the problem. We assume the world lines begin at the same point (the left picture), then cross once again and tend towards their asymptotic forms (the right picture). This type of evolution is typical of sources and detectors that move with opposite accelerations. The world-lines $r \mapsto x^\mu(r)$ and  $s \mapsto y^\mu(s)$ represent, respectively, the source and the detector. The auxiliary world-line $q \mapsto z^\mu(q)$ represents the rest frame. The points $x^\mu(r)$, $z^\mu(q)$, and $y^\mu(s)$ are located on the same light cone. The same concerns their versions shifted by $\Delta r$,  $\Delta q$,  and $\Delta s$. In effect, the three proper-time parameters are not independent of one another.  In order to find the explicit form of $s(r)$ it is simplest to split the derivation into two steps, $q(r)$ and $s(q)$. }
\label{Fig1}
\end{figure}

Assume an unstable system propagating along a world-line $r\mapsto x^\mu(r)$ in Minkowski space decays exponentially in its comoving  reference frame  \cite{Rossi1941,Frisch1963}. The survival probability  at proper time $r$  is given by $p(r)=e^{-r/r_0}$. Here, $r\ge 0$ is the proper time computed along the world-line and $r_0$ is the mean lifetime  (we measure proper time in units of length). Now, consider another world-line, $s \mapsto y^\mu(s)$, describing a detector which absorbs at $y^\mu(s)$ the light signal emitted 
at $x^\mu(r)$. Here, $s\ge 0$ is the proper time of the detector, computed along the detector's world-line. The difference $y^\mu(s)-x^\mu(r)$ is a future-pointing null vector. We also assume that at the moment the decay begins, at $r=s=0$, both objects are at the same point in space-time, $x^\mu(0)=y^\mu(0)$. If the decay takes place at proper time $r$, it will be detected at proper time $s=s(r)$ which depends on both world-lines (Fig.~\ref{Fig1}).

The problem is nontrivial in that accelerated sources or detectors may lead to event horizons. In particular, a particle that decays exponentially in its rest frame may approach the event horizon of the detector in a finite time $r_{\rm max}$, but the detector will have to wait forever to see the source cross the horizon if  $s(r_{\rm max})=\infty$. Accordingly, the survival probability of the detector, $P(s)$, cannot be exponential because  $P(\infty)\neq 0$.

Assuming an ideal detector,  the  probability of non-detection equals the probability of non-emission, provided signal retardation and time dilation in the detector are taken into account:
\be
P(s(r))=p(r), \quad P(s)=p(r(s)). \label{Psr}
\ee
The question is what are the forms of $s(r)$, $r(s)$ and $P(s)$? 

The observed $P(s)$ cannot be exponential if an event horizon occurs. However, we will show below that $P(s)$ is never exponential when there are accelerations, regardless of the presence or absence of a horizon.
We find an explicit closed-form solution for the case of sources and detectors moving with constant (but opposite) accelerations,
\be
P(s)&=&
\left\{
\begin{array}{rl}
\left(1- \frac{|a|}{|b|} c^{-2}(U_0-|\bm U|)(V_0-|\bm V|)(e^{|b|s/c^2}-1)\right)^{\frac{c^2}{|a|r_0}},&  0\le s\le s_1,\\
\left(1+\frac{|a|}{|b|}c^{-2}(U_0+|\bm U|)(V_0+|\bm V|)(1-e^{-|b|s/c^2})\right)^{-\frac{c^2}{|a|r_0}}, & s\ge s_1,
\end{array}
\right.\label{P2}
\ee
where $s_1=s(r_1)$ is the value of the detector proper time corresponding to the moment the world-lines cross the second time (that is, if  $x^\mu(r_1)=y^\mu(s_1)$ for a nonzero value $r=r_1>0$; the first time they cross occurs at the initial condition $r=s=0$). Parameters $a$ and $b$ are the accelerations of the source and the detector, respectively, whereas $U^\mu$ and $V^\mu$ are their  initial world-velocities. For a nonzero detector acceleration, $b\neq 0$, one finds 
\be
P(\infty)
&=&
\left(1+\frac{|a|}{|b|}c^{-2}(U_0+|\bm U|)(V_0+|\bm V|)\right)^{-\frac{c^2}{|a|r_0}}
=e^{-r_{\rm max}/r_0}\neq 0.
\label{P3}
\ee
A finite value of $r_{\rm max}$ means that the event horizon occurs for the detector.
The limit $b\to 0$ describes a detector that moves with no acceleration, 
\be
\lim_{b\to 0} P(s) 
&=& 
\left\{
\begin{array}{rl}
\left(1- |a|(U_0-|\bm U|)(V_0-|\bm V|)s/c^4\right)^{\frac{c^2}{|a|r_0}},&  0\le s\le s_1,\\
\left(1+ |a|(U_0+|\bm U|)(V_0+|\bm V|)s/c^4\right)^{-\frac{c^2}{|a|r_0}}, & s\ge s_1.
\end{array}
\right.\label{P3}
\ee
The exponential decay is then replaced by an exact power law (of a Zipf-Mandelbrot type), but  the event horizon disappears.

At the other extreme is the case of an accelerating detector and a non-accelerating source,
\be
\lim_{a\to 0} P(s) 
&=& 
\left\{
\begin{array}{rl}
\exp\left(- \frac{1}{|b|r_0} (U_0-|\bm U|)(V_0-|\bm V|)(e^{|b|s/c^2}-1)\right),&  0\le s\le s_1,\\
\exp\left(-\frac{1}{|b|r_0}(U_0+|\bm U|)(V_0+|\bm V|)(1-e^{-|b|s/c^2})\right), & s\ge s_1,
\end{array}
\right.\label{P4}
\ee
which is a Gumbel-type double-exponential distribution.

Formulas (\ref{P3}) and (\ref{P4}) show that the two limits commute, leading to a Doppler-type formula, 
\be
\lim_{a\to 0}\lim_{b\to 0} P(s) 
&=&
\lim_{b\to 0}\lim_{a\to 0} P(s) \\
&=&
\exp\left(- \frac{s}{r_0} \frac{U_0-|\bm U|}{V_0+|\bm V|}\right)
=
\exp\left(- \frac{s}{r_0} \frac{V_0-|\bm V|}{U_0+|\bm U|}\right),\label{P5b}
\ee
for any $s\ge 0$ (since $s_1\to\infty$).

The discussed deformation of the exponential law is a purely kinematic effect caused by the relativistic time dilation in the detector and the finite speed of signal propagation. It should not be confused with other possible reasons for deviations from a simple exponential formula found in experiments. 

For example, we know that spontaneous decays of unstable quantum states are only approximately exponential \cite{Khalfin,Levy,Newton,Winter,Onley} (see \cite{Wang} for a recent review). Moreover, even classical accelerated systems are subject to additional forces and potentials that can affect their lifetime, as any pyrotechnist knows well. In quantum field theory models of particle decay, acceleration leads to additional phenomena of the Fulling-Davies-Unruh type 
\cite{F,D,U,Unruh,AM1994,Mueller1997,Matsas,Matsas2,Blasone,Luciano}. However, all these effects refer to the level of $p(r)$, while what we are discussing is a result of a non-trivial $s(r)$. The latter is not affected by the form $p(r)$, but by the relation between $x^\mu(r)$ and $y^\mu(s)$.

The formula we derive below will have its analogue for any form of the decay, and the result will be again given by (\ref{Psr}). Our derivation is generally valid for basically any motion of sources and detectors. Systems accelerating in more complicated ways will lead to more complicated forms of $s(r)$, still satisfying our differential equation for $ds(r)/dr$ (see, however, the remarks in Sec.~\ref{Sec3} and Sec.~\ref{Sec6} on the role of the $t$-$z$ hyperplane). 

Somewhat unexpectedly, the probability given by (\ref{P2}) (especially when written as (\ref{final P})) turns out to have exactly the form postulated over two decades ago by Tsallis, Bemski and Mendes \cite{TBM} in their analysis of ligand binding dynamics \cite{Austin1974,Austin1975}. Moreover, inspired by \cite{TBM},  Montemurro \cite{Montemurro} showed that the same probability function can be used to adequately model observable deviations from the Zipf-Mandelbrot law in quantitative linguistics. All these links with the problem discussed in the present paper were completely unexpected and hard to anticipate. It should be stressed, though, that neither \cite{TBM} nor \cite{Montemurro} were capable of deriving  (\ref{P2})   and (\ref{final P}) from first principles --- rather, it was an educated guess based on the structure of the data and an ad hoc modification of some differential equations. 

A step towards a first-principles derivation of (\ref{P2}) and (\ref{final P}), although completely unrelated to what we discuss in the present paper, was made in \cite{CN}. What we showed in \cite{CN} was  that  probabilities of the form (\ref{final P}) automatically occur in thermodynamics based on R\'enyi's entropy, provided one seriously treats the original R\'enyi's construction. Namely, R\'enyi in his very first and rarely quoted paper \cite{Renyi} replaced the linear average $\sum_p p\log(1/p)$ of the Shannon random variable $\log (1/p)$ by its Kolmogorov-Nagumo average $\varphi^{-1}\Big(\sum_p p\,\varphi\big(\log(1/p)\big)\Big)$ \cite{KN1,KN2}, and this led him to the  well-known formula $\frac{1}{1-\alpha}\log\sum_p p^\alpha$ for the $\alpha$-entropy \cite{Jizba1,Jizba2}. The conclusion of \cite{CN} was that (\ref{final P}) is a consequence of applying the same Kolmogorov-Nagumo averaging to both 
$\log (1/p)$ and the constraints typical of maximum entropy principles. Although the exact form (\ref{final P}) occurred for the concrete choice of $\varphi$ made by R\'enyi,  the formalism of \cite{CN} allowed for further generalizations discussed in detail by Naudts in his monograph \cite{N}. A similar degree of generality is encountered in the formalism introduced in the present paper if one allows for more general forms of accelerated motion.

\begin{figure}
\includegraphics[width=8 cm]{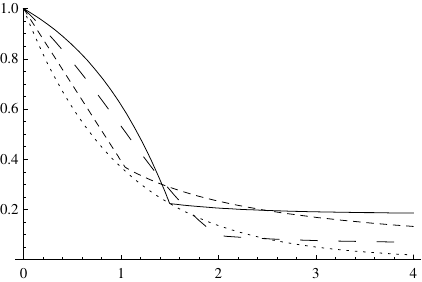}
\caption{Probability $P(s)$ given by (\ref{P2}). The average lifetime parameter   $r_0=1$ (in arbitrary units); the other parameters correspond to $P(s)$ written as in (\ref{final P}). The ordinary exponential decay $P(s)=e^{-s/r_0}$ is found for $A=B=0$, that is when both the source and the detector remain in the same inertial reference frame (actually, the dotted line shows the case of $A=B=10^{-6}$, $R=0.5$, $S=1$, which is here indistinguishable from $e^{-s}$). The short-dashed line represents the decay (\ref{P3}) with $A=-1$, $R=0.5$ as seen by a detector moving with constant velocity. The long-dashed line is the decay as observed by an accelerated detector, when the source moves with constant velocity, here with  $A=10^{-6}$, $B=1$, $R=0.5$, $S=1$. The full line represents $P(s)$  with $A=-1$, $B=1$, $R=0.5$, $S=1$, the case of accelerated sources and detectors.} 
\label{Fig7}
\end{figure}

However, the methods introduced in the present article are purely relativistic and have nothing to do with thermodynamics. 

We begin in Section~\ref{Sec3} with determining the form of the map $r\mapsto s(r)$ which relates the proper time of emission with the one of the detection.  We derive a differential equation satisfied by $s(r)$ once the explicit forms of $x^\mu(r)$ and $y^\mu(s)$ are given. In Section~\ref{Sec4} we restrict our analysis to the concrete case of sources and detectors that move with constant accelerations. The two world-lines cross twice: at the initial condition and then once again, ultimately separating into their asymptotic forms at infinity. The two stages, between the crossing points and behind the second of them, are qualitatively different and have to be treated independently. Once we have obtained the explicit form of $s(r)$ we are able to write our final formula for $P(s)$, the task completed in Section~\ref{Sec5}. Finally, in Section~\ref{Sec6} we briefly discuss implications of our analysis for the problem of clock synchronization in neighborhoods of event horizons.

\section{Relation between proper times}
\label{Sec3}

\begin{figure}
\includegraphics[width=5.5 cm]{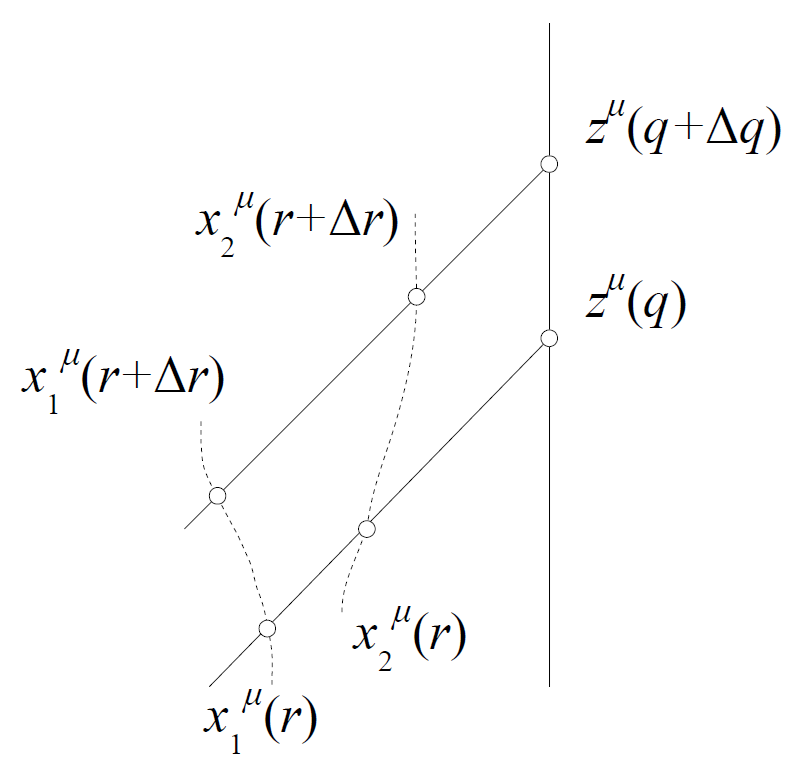}
\includegraphics[width=6 cm]{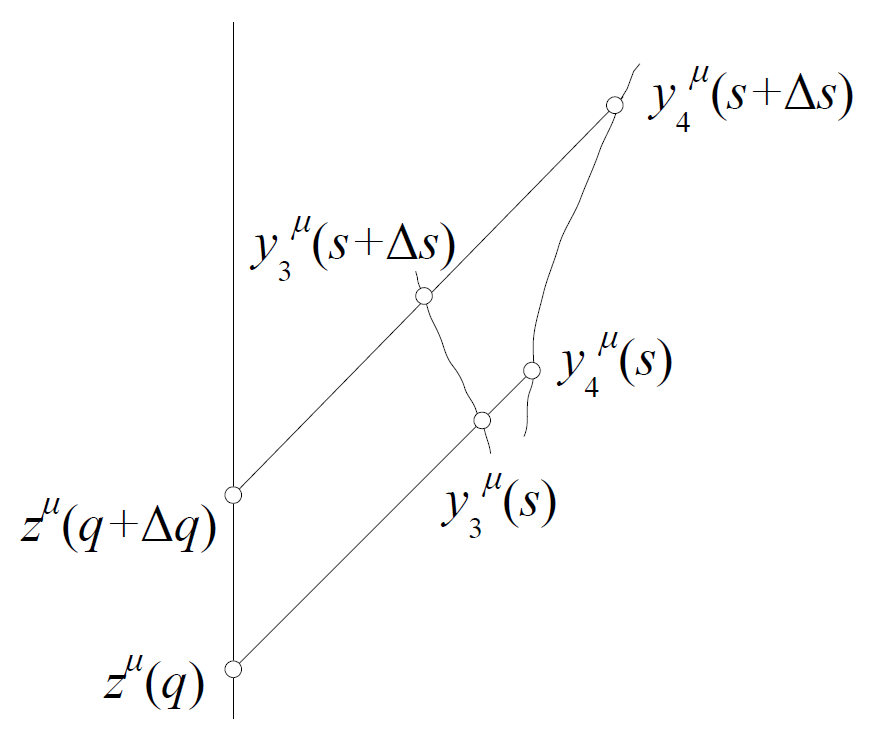}
\caption{Splitting the source-detector relation into two steps (the stage between the two crossing points; the stage behind the second crossing point will look analogously, see Fig.~\ref{Fig3b}). Left: Two types of relations between the world-lines of the source, $x_j^\mu(r)$, $j=1,2$, and those of the rest-frame detectors, $z^\mu(q)$ (the source either moves toward the detector or escapes from it). Right: Similar relations between the world-lines of the rest-frame source, $z^\mu(q)$, and those of the detectors, $y_k^\mu(s)$, $k=3,4$. Events $x_j^\mu$, $z^\mu$, and  $y_k^\mu$ are causally related by null vectors. } 
\label{Fig2}
\end{figure}

Assume the three world-lines shown in Fig.~\ref{Fig1} are restricted to some $t$-$z$ plane of the Minkowski space,
\be
r &\mapsto& x^\mu(r)=\big( x^0(r),0,0,x^3(r)\big),\label{1}\\
s &\mapsto& y^\mu(s)=\big( y^0(s),0,0,y^3(s)\big),\label{2}\\
q &\mapsto& z^\mu(q)=\big( q,0,0,0\big).\label{3}
\ee
(\ref{1})  and (\ref{2}) describe, respectively, the source and the detector, whereas (\ref{3}) is an auxiliary world-line introduced just for convenience and describes the origin $\bm z(q)=(0,0,0)$ of the rest frame.

The metric has signature $(+,-,-,-)$ and we in general simplify notation by skipping the two vanishing components. 
The parametrizations are given in terms of proper times,
\be
x^\mu(r+\Delta r)
&=&
x^\mu(r)+\Delta x^\mu(r),\\
y^\mu(s+\Delta s)
&=&
y^\mu(s)+\Delta y^\mu(s),\\
\dot x_\mu(r)\dot x^\mu(r)
&=&
\dot y_\mu(s)\dot y^\mu(s)=\dot z_\mu(q)\dot z^\mu(q)=1.
\ee
The dots represent proper-time derivatives. The world-vectors $y^\mu(s)-z^\mu(q)$, $z^\mu(q)-x^\mu(r)$, $y^\mu(s+\Delta s)-z^\mu(q+\Delta q)$, and $z^\mu(q+\Delta q)-x^\mu(r+\Delta r)$ are null. 

\begin{figure}
\includegraphics[width=8 cm]{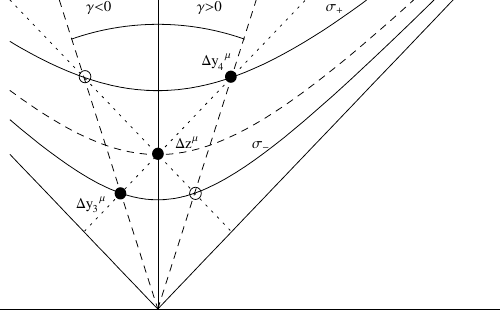}
\includegraphics[width=8 cm]{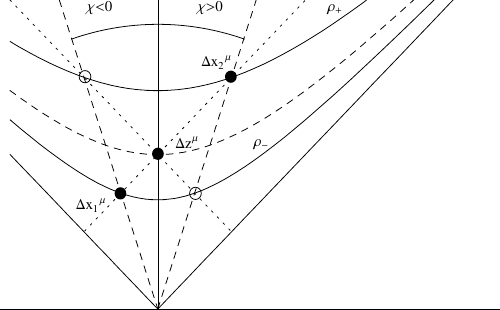}
\caption{
The geometry between the two crossing points.
Left: The geometry of the quadratic equation (\ref{15}) and its two solutions $\sigma_\pm$ given by  (\ref{16}). The fact that $\sigma_+\ge \sigma_-$ implies $\sigma_+^2=\Delta y_{4\mu}(s)\Delta y_{4}^\mu(s)$, $\sigma_-^2=\Delta y_{3\mu}(s)\Delta y_{3}^\mu(s)$.
Right: The geometry of the quadratic equation (\ref{15'}) and its two solutions $\rho_\pm$ given by   (\ref{16'}).  The fact that $\rho_+\ge \rho_-$ implies  $\rho_+^2=\Delta x_{2\mu}(r)\Delta x_{2}^\mu(r)$, $\rho_-^2=\Delta x_{1\mu}(r)\Delta x_{1}^\mu(r)$. }
\label{Fig3}
\end{figure}

\begin{figure}
\includegraphics[width=8 cm]{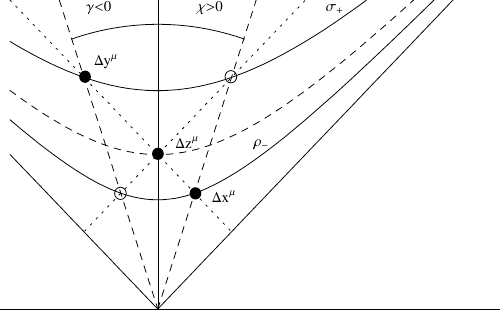}
\caption{
Behind the second crossing point one finds 
$\sigma_+^2=\Delta y_{\mu}(s)\Delta y^\mu(s)$, $\rho_-^2=\Delta x_{\mu}(r)\Delta x^\mu(r)$,  $\gamma(s)<0$, and $\chi(r)>0$.}
\label{Fig3b}
\end{figure}

We begin with establishing a relation between
$\sigma^2=\Delta y_\mu(s)\Delta y^\mu(s)$, $\rho^2=\Delta x_\mu(r)\Delta x^\mu(r)$, and $\Delta q$.
The  null (i.e. light-like) vectors  in Fig.~\ref{Fig2} are parallel, in consequence of restricting the motions to the $t$-$z$ hyperplane in the Minkowski space (light-like world-lines in $(1+1)$-dimensional Minkowski  $t$-$z$  space are given by $ct=\pm z+\textrm{const}$, and all such lines are either parallel or perpendicular). In particular,
\be
y^\mu(s+\Delta s)-z^\mu(q+\Delta q)
&=&
y^\mu(s)+\Delta y^\mu(s)-z^\mu(q)-\Delta z^\mu(q)
\sim y^\mu(s)-z^\mu(q),\\
z^\mu(q+\Delta q)-x^\mu(r+\Delta r)
&=&
z^\mu(q)+\Delta z^\mu(q)-x^\mu(r)-\Delta x^\mu(r)
\sim
z^\mu(q)-x^\mu(r),
\ee
so that
\be
\Delta y^\mu(s)-\Delta z^\mu(q)
&\sim& y^\mu(s)-z^\mu(q),\\
\Delta z^\mu(q)-\Delta x^\mu(r)
&\sim&
z^\mu(q)-x^\mu(r),
\ee
a fact implying that 
$\Delta y^\mu(s)-\Delta z^\mu(q)$ are $\Delta z^\mu(q)-\Delta x^\mu(r)$  are null as well. On the other hand, $\Delta x^\mu(r)$, $\Delta y^\mu(s)$, and $\Delta z^\mu(q)$ are timelike and future-pointing, so their relations can be parametrized by a hyperbolic coordinate,
\be
\Delta z_\mu(q)\Delta y^\mu(s)
&=&\sigma \Delta q \cosh\gamma,\\
\Delta z_\mu(q)\Delta x^\mu(r)
&=& \rho \Delta q\cosh\chi.
\ee
Since $\Delta y^\mu(s)-\Delta z^\mu(q)$ and $\Delta z^\mu(q)-\Delta x^\mu(r)$ are null, we find
\be
(\sigma/\Delta q)^2-2 (\sigma/\Delta q)\cosh\gamma+1 &=& 0,\label{15}\\
(\rho/\Delta q)^2-2 (\rho/\Delta q)\cosh\chi+1 &=& 0.\label{15'}
\ee
Accordingly,
\be
\sigma_\pm/\Delta q
&=&\cosh\gamma \pm|\sinh\gamma|
=
e^{\pm|\gamma|},
\label{16}
\\
\rho_\pm/\Delta q
&=&\cosh\chi \pm|\sinh\chi|
=
e^{\pm|\chi|}.
\label{16'}
\ee
In the limit $\Delta q\to 0$ we find two differential equations that link the three proper times,
\be
\frac{ds(q)}{dq} &=&
e^{\pm |\gamma(s(q))|},\label{25}\\
\frac{dr(q)}{dq} &=&
e^{\pm |\chi(r(q))|}\label{26}.
\ee
Fig.~\ref{Fig3} and Fig.~\ref{Fig3b} explain how to relate the signs in (\ref{25})--(\ref{26}) with the signs of $\gamma(s(q))$ and $\chi(r(q))$. First of all, $\sigma_+\ge \sigma_-$ and $\rho_+\ge\rho_-$. For the stage between the crossing points this implies $\sigma_+^2=\Delta y_{4\mu}(s)\Delta y_{4}^\mu(s)$ and $\rho_+^2=\Delta x_{2\mu}(r)\Delta x_{2}^\mu(r)$. Secondly,  the hyperbolic parameters are positive if the particle moves to the right, the case of the parts of the world-lines denoted in Fig.~\ref{Fig3} by $\Delta y_{4}^\mu(s)$ and $\Delta x_{2}^\mu(r)$, and in Fig.~\ref{Fig2} by $y_{4}^\mu(s)$ and $x_{2}^\mu(r)$. The same reasoning explains why the signs in $\sigma_\pm$ and $\rho_\pm$ are negative for $y_{3}^\mu(s)$ and $x_{1}^\mu(r)$, that is when the particles move to the left and thus the hyperbolic parameters are negative. Collecting all these cases we conclude that for trajectories whose shape is depicted in the left part of Fig.~\ref{Fig1}, the equations to solve are
\be
\frac{ds(q)}{dq} &=&
e^{\gamma(s(q))},\label{25'}\\
\frac{dr(q)}{dq} &=&
e^{\chi(r(q))}\label{26'}.
\ee
The chain rule for derivatives thus implies
\be
\frac{ds(q(r))}{dr} &=&
e^{\gamma(s(q(r)))-\chi(r)}\label{27}.
\ee
Denoting $r=s_a$, $s=s_b$, $\gamma(s)=\gamma_a(s_a)$, $\chi(r)=\gamma_b(s_b)$, we can rewrite (\ref{27}) as
\be
\frac{ds_a}{ds_b} &=&
e^{\gamma_a(s_a)-\gamma_b(s_b)}=\beta{_a}{_b}\label{27''},
\ee
exhibiting the cocycle property,
\be
\beta{_a}{_b}\beta{_b}{_c}=\beta{_a}{_c},\label{coc}
\ee
of the world-lines and their proper-time parametrizations, i.e.
the auxiliary rest-frame world-line in Fig.~\ref{Fig1} can be replaced by any world-line passing through the two crossing points.


An analogous analysis applies to the section behind the second crossing point (Fig.~\ref{Fig3b}),  where we find that $\gamma(s)$ is negative, $\gamma(s)=-|\gamma(s)|$ but the sign in $\sigma_\pm$ is positive. Hence,
\be
\frac{ds(q)}{dq} &=&
e^{|\gamma(s(q))|}=e^{-\gamma(s(q))}\label{25''}.
\ee
On the other hand, $\chi(r)$ is here positive, but the sign in $\rho_\pm$ is negative, hence
\be
\frac{dr(q)}{dq} &=&
e^{-|\chi(r(q))|}=e^{-\chi(r(q))}\label{26''}.
\ee 
The chain rule for derivatives now implies
\be
\frac{ds(q(r))}{dr} &=&
e^{-\gamma(s(q(r)))+\chi(r)}\label{27'}.
\ee
(\ref{27'}) possess the same cocycle property (\ref{coc}) as the world-lines connecting the two crossing points.

In the next Section we restrict the analysis to the particular case of world-lines  occurring for sources and detectors moving with constant accelerations.


\section{Uniformly accelerated sources and detectors}
\label{Sec4}

\begin{figure}
\includegraphics[width=3 cm]{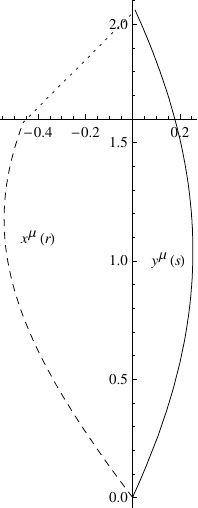}
\caption{
World-lines $x^\mu(r)$ (dashed) and  $ y^\mu(s(r))$ (full) given by (\ref{x})--(\ref{gammas}), yet before the second crossing point,   for $A=-1$, $R=0.5$, $B=1$, $S=1$, and $0\le r\le 1.4$ (in arbitrary units). The dotted line is on the light cone joining the endpoints of $x^\mu(r)$  and  $ y^\mu(s(r))$. This is the world-line of the light signal emitted at $x^\mu(r)$  and  detected at $ y^\mu(s(r))$.}
\label{Fig4}
\end{figure}

A particle that propagates along the world-line $t\mapsto (ct,0,0,z(t))$ with constant acceleration $a$,   satisfies the Newton equation
\be
\frac{d}{dt}\frac{v(t)}{\sqrt{1-\frac{v(t)^2}{c^2}}}
=
a,
\quad
v(t)
=
\frac{dz(t)}{dt}.
\ee
Its solution is given by
\be
v(t)
&=&
c\frac{at+U}{\sqrt{c^2 +(at+U)^2}},
\\
z(t)
&=&
z(0)+\frac{c}{a}\sqrt{c^2 +(at+U)^2}-\frac{c}{a}\sqrt{c^2 +U^2}.
\ee
Here, $U$ is the initial world-velocity, $\bm U(0)=(0,0,U)$. 
The curve $t\mapsto \big(ct,z(t)\big)$  is a hyperbola parametrized by time $t$. The integrated proper time, 
$c^2d\tau^2=c^2 dt^2-dz^2$, $\tau(0)=0$, when evaluated along the world-line, yields
\be
c\tau(t)
&=&
(c^2/a)\arc\sinh(at/c+U/c)-(c^2/a)\arc\sinh(U/c ) .
\ee
Assuming $(ct(0),z(0))=(0,0)$, we find
\be
\big(ct(\tau),z(\tau)\big)
&=&
\frac{c^2}{a}\Big(\sinh\big(a\tau/c+\arc\sinh(U/c)  \big)
-
U/c,
\cosh\big(a\tau/c+\arc\sinh(U/c)  \big)-\sqrt{1 +(U/c)^2}
\Big).
\label{w-line}
\ee
World-line (\ref{w-line}) will play in our analysis the role of either $x^\mu(r)$ or $y^\mu(s)$ (we assume the source moves with positive constant acceleration, $a=\textrm{const}>0$, and initially negative world-velocity 
$U<0$;  the detector acceleration is constant and negative, $b=\textrm{const}<0$, but its initial world-velocity is positive, $V>0$). To this end, let us write the two world-lines as follows (Fig.~\ref{Fig4}),
\be
\big(x^0(r),x^3(r)\big)
&=&
\frac{1}{A}
\big(\sinh(AR)-\sinh(-Ar +AR),\cosh(AR)-\cosh(-Ar +AR)\big),\label{x}\\
\big(\dot x^0(r),\dot x^3(r)\big)
&=&
\big(\cosh(-Ar +AR),\sinh(-Ar +AR)\big)=\big(\cosh\chi(r),\sinh\chi(r)\big),\\
A &=& -a/c^2=-|A|,\quad a> 0,\\
R &=& -(c^2/a)\arc\sinh(U/c)=|R|,\quad U< 0,\\
\chi(r) &=& -Ar +AR=|A|r-|A|R,\label{chir}\\
\big(y^0(s),y^3(s)\big)
&=&
\frac{1}{B}
\big(\sinh(BS)-\sinh(-Bs +BS),\cosh(BS)-\cosh(-Bs +BS)\big),\label{y}\\
\big(\dot y^0(s),\dot y^3(s)\big)
&=&
\big(\cosh(-Bs +BS),\sinh(-Bs +BS)\big)=\big(\cosh\gamma(s),\sinh\gamma(s)\big),\\
B &=& -b/c^2=|B|,\quad b< 0,\\
S &=& -(c^2/b)\arc\sinh(V/c)=|S|,\quad V>0,\\
\gamma(s)
&=&
-Bs+BS=-|B|s+|B|S.
\label{gammas}
\ee
Expressions such as $ |A|R$ and  $|B|S$ are independent of accelerations.

\subsection{Between the crossing points}

The analysis given in the previous Section can be now directly applied to (\ref{x})--(\ref{gammas}). Equations 
\be
\frac{ds(q)}{dq} &=&
e^{\gamma(s(q))}=e^{-|B|s(q)+|B|S},\quad s(0)=0,\\
\frac{dr(q)}{dq} &=&
e^{\chi(r(q))}=e^{|A|r(q)-|A|R},\quad r(0)=0,
\ee
imply
\be
r(s)
&=&
\ln\left(1- \frac{|A|}{|B|} e^{-|A|R-|B|S}(e^{|B|s}-1)\right)^{-\frac{1}{|A|}},\label{rs}\\
s(r)&=&
\ln\left(1+\frac{|B|}{|A|}e^{|A|R+|B|S}(1-e^{-|A|r})\right)^{\frac{1}{|B|}}.\label{sr}
\ee
The solution can be directly cross-checked by
\be
\frac{ds(r)}{dr} &=&
e^{\gamma(s(r))-\chi(r)},
\ee
with $\chi(r)$, $\gamma(s)$, and $s(r)$ given  by (\ref{chir}), (\ref{gammas}),   and (\ref{sr}), respectively.

\subsection{Behind the second crossing point}

The second crossing point is determined by
\be
\big(x^0(r),x^3(r)\big)
&=&
\big(y^0(s(r)),y^3(s(r))\big), \quad r>0,
\ee
where $s(r)$ is the solution (\ref{sr}). It occurs for $r=r_1$,
\be
r_1
&=&
\frac{1}{|A|}\ln\frac{|A|e^{|A|R+|B|S}+|B| }{|A|e^{-|A|R-|B|S}+|B| },\label{cross r}\\
s(r_1)
&=&
\frac{1}{|B|}\ln\frac{|A|+|B| e^{|A|R+|B|S} }{|A|+|B|e^{-|A|R-|B|S} }=s_1,\label{cross s}
\ee
and is located at 
\be
x^0(r_1)
&=&
y^0(s(r_1))\\
&=&
2\sinh(|A|R+|B|S)\frac{|A|\cosh{|B|S}+|B|\cosh{|A|R}}{(|A| e^{|A|R+|B|S}+|B| )(|A|e^{-|A|R-|B|S}+|B|) }\label{cross 0}
,\\
x^1(r_1)
&=&
y^1(s(r_1))\\
&=&
2\sinh(|A|R+|B|S)
\frac{|A|\sinh{|B|S}-|B| \sinh{|A|R}}{(|A| e^{|A|R+|B|S}+|B| )(|A|e^{-|A|R-|B|S}+|B|) }.\label{cross 1}
\ee
Behind the second crossing point we have to solve
\be
\frac{ds(q)}{dq} &=&
e^{-\gamma(s(q))}=e^{|B|s(q)-|B|S},\\
\frac{dr(q)}{dq} &=&
e^{-\chi(r(q))}=e^{-|A|r(q)+|A|R},
\ee
which is equivalent to
\be
\frac{ds(r)}{dr} &=&
e^{-\gamma(s(r))+\chi(r)}.
\ee
Collecting the above two cases, 
one finds that
\be
s(r)
&=&
\left\{
\begin{array}{rl}
\frac{1}{|B|}\ln\left(1+\frac{|B|}{|A|}e^{|A|R+|B|S}(1-e^{-|A|r})\right), & 0\le r\le r_1,\label{sr'}\\
-\frac{1}{|B|}
\ln\left(1-\frac{|B|}{|A|}e^{-|A|R-|B|S}(e^{|A|r}-1)\right), &  r_1\le r<r_{\rm max},
\end{array}
\right.
\ee
is a solution to
\be
\frac{ds(r)}{dr}
&=&
\left\{
\begin{array}{rl}
e^{\gamma(s(r))-\chi(r)}, & 0\le r\le r_1,\label{eqsr}\\
e^{-\gamma(s(r))+\chi(r)}, &  r_1\le r<r_{\rm max}.
\end{array}
\right.
\label{55}
\ee
In (\ref{sr'}) the case $0\le r\le r_1$ corresponds to (\ref{sr}). $s(r)$ is continuous at $r_1$ but its derivative (\ref{55})  is not.
Let us note that  $r$ is limited by its maximal value
\be
r_{\rm max} 
&=&\frac{1}{|A|}\ln \left(1+\frac{|A|}{|B|}e^{|A|R+|B|S}\right)
\ee
with $s(r_{\rm max})=\infty$. At the event horizon, i.e. for  $r\ge r_{\rm max}$, the source  becomes invisible to the detector, but
\be
\lim_{B\to 0}r_{\rm max}=\infty.
\ee
A detector at rest can monitor the accelerated system forever.

The inverse map, $r(s)$, reads
\be
r(s)
&=&
\left\{
\begin{array}{rl}
-\frac{1}{|A|}\ln\left(1- \frac{|A|}{|B|} e^{-|A|R-|B|S}(e^{|B|s}-1)\right), & 0\le s\le s_1,\\
\frac{1}{|A|}
\ln
\left(
1+\frac{|A|}{|B|}e^{|A|R+|B|S}(1-e^{-|B|s})
\right), &  s_1\le s,
\end{array}
\right.
\label{full r}
\ee
with $s_1=s(r_1)$ given by  (\ref{cross s}). 
In particular,
$r(\infty)=r_{\rm max}$,
as expected.
\begin{figure}
\includegraphics[width=8 cm]{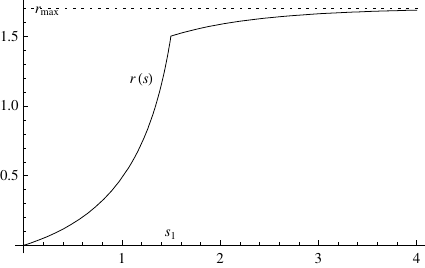}
\caption{Function $r(s)$ given by (\ref{full r}) for the same parameters as in Fig.~\ref{Fig4}. The point of second crossing, $s=s_1$, is visible as the point of non-differentiability of $r(s)$. The dotted line is the asymptotic value $r_{\rm max}=r(\infty)$ occurring at the event horizon.} 
\label{Fig6}
\end{figure}

\section{Effective non-exponential decay}
\label{Sec5}

We are ready to write the final formula for the probability measured by the detector,
\be
 P(s) &=& e^{-r(s)/r_0}
\\
&=&
\left\{
\begin{array}{rl}
\left(1- \frac{|A|}{|B|} e^{-|A|R-|B|S}(e^{|B|s}-1)\right)^{\frac{1}{|A|r_0}},&  0\le s\le s_1,\\
\left(1+\frac{|A|}{|B|}e^{|A|R+|B|S}(1-e^{-|B|s})\right)^{-\frac{1}{|A|r_0}}, & s\ge s_1
\end{array}
\right.
\label{final P}
\ee
with
\be
s_1
=
\frac{1}{|B|}\ln\frac{|A|+|B| e^{|A|R+|B|S} }{|A|+|B|e^{-|A|R-|B|S} },
\ee
and
\be
P(\infty)=e^{-r_{\rm max}/r_0}.
\ee
Formulas (\ref{x})--(\ref{gammas}) imply $|A|=|a|/c^2$, $|B|=|b|/c^2$,
\be
e^{\pm|A||R|}
&=&
(U_0\pm |U|)/c,\\
e^{\pm|B||S|}
&=&
(V_0\pm|V|)/c,
\ee
where $(U_0,0,0,U)$ and $(V_0,0,0,V)$ are the initial world-velocities of the source and the detector, respectively.
This ends the derivation of formula (\ref{P2}).

We can directly compare (\ref{final P}) with the probability postulated in \cite{TBM}, 
\be
P(s)
&=&
\left(
1+\frac{\lambda}{\mu}\left(e^{(q-1)\mu s}-1\right)
\right)^{-\frac{1}{q-1}},
\label{68}
\ee
in order to fit the molecular data from \cite{Austin1974,Austin1975}, and employed  in \cite{Montemurro} in fitting the Zipf-law data from 36 plays by Shakespeare and 56 books by Dickens.
Although there is no reason to believe that there are any links between the statistics of muon decays  and frequencies of words used by Shakespeare, it is nevertheless intriguing that the same non-evident functional dependence on parameters is found.

In the limit $B\to0 $, corresponding to the detector moving with constant velocity, we obtain a power law,
\be
\lim_{B\to 0} P(s) 
&=& 
\left\{
\begin{array}{rl}
\left(1- |A|e^{-|A|R}e^{-|B|S}s\right)^{\frac{1}{|A|r_0}},&  0\le s\le s_1,\\
\left(1+|A|e^{|A|R}e^{|B|S}s\right)^{-\frac{1}{|A|r_0}}, & s\ge s_1
\end{array}
\right.
\label{final P B=0}\\
&=&
\left\{
\begin{array}{rl}
\left(1- |a|(U_0-|U|)(V_0-|V|)s/c^4\right)^{\frac{c^2}{|a|r_0}},&  0\le s\le s_1,\\
\left(1+ |a|(U_0+|U|)(V_0+|V|)s/c^4\right)^{-\frac{c^2}{|a|r_0}}, & s\ge s_1.
\end{array}
\right.
\ee
with
\be
s_1 =
2\frac{\sinh(|A|R+|B|S)}{|A|}
=2\frac{
U_0|V|
+
|U|V_0
}{|a|}.
\ee

Fig.~\ref{Fig7} shows $P(s)$ for $r_0=1$ and various values of the remaining parameters. 

\section{Final remarks}
\label{Sec6}

The effect discussed  is purely kinematic. We have focused on the exponential case and uniform accelerations because of the simplicity of the resulting formulas, but our results apply in principle to any decay and any motion of the sources and detectors. For example, considering various quantum and field theoretical phenomena that occur as a result of accelerations \cite{AM1994,Mueller1997,Matsas,Matsas2,Blasone,Luciano}, one expects modifications of $p(r)=e^{-r/r_0}$, but in any case the final result will be given by $P(s)=p(r(s))$, provided of course that all signals emitted at $x^\mu(r)$ are detected at $y^\mu(s)$.

Among the assumptions that simplify the argument is that $x^\mu(r)$ and $y^\mu(s)$ are confined to the same $t$-$z$ hyperplane for all $r$ and $s$. Source and detector thus move along the same straight line in space, making the problem effectively one-dimensional. If one relaxes this condition and considers more general motions, it would no longer be true that $y^\mu(s)-x^\mu(r)$ and $y^\mu(s+\Delta s)-x^\mu(r+\Delta r)$ are parallel for any $\Delta s$ and $\Delta r$. The simplicity of the argument would be lost. In this sense, our formulas are not the most general.

Finally, it should be mentioned that Fig.~\ref{Fig4} suggests yet another perspective on the subject of the paper. Namely, let us note that both $r$ and $s(r)$ are proper times measured by clocks propagating along the two world-lines. $s(r)$ is the proper time registered by the detector at the moment it detects (i.e. {\it observes\/}) the light signal emitted at proper time $r$ (as measured by the clock of the source).  An observer propagating along $s\mapsto y^\mu(s)$  sees at his proper time $s(r)$ the time $r$ as it appears on  the clock of the source located at $x^\mu(r)$. The relation between the time $r$ of emission and the time $s(r)$ of observation is one-to-one. The equivalence $r\leftrightarrow s$ may be regarded as a procedure of clock synchronization in a neighborhood of an event horizon, a problem of great importance for quantum cryptography, quantum teleportation, and --- more generally --- the studies of entanglement in non-trivial space-times \cite{Mann}.

\acknowledgments

Calculations were carried out at the Academic Computer Center in Gda{\'n}sk. The work was supported by the CI TASK grant `Non-Newtonian calculus with interdisciplinary applications'. I'm indebted to Kamil Nalikowski and Pasquale Cirillo for inspiring discussions.  Special thanks to Jan Naudts for years of collaboration on generalized statistics.

\end{document}